\documentclass[sigconf, screen, nonacm]{acmart}
\AtBeginDocument{%
  \providecommand\BibTeX{{%
    \normalfont B\kern-0.5em{\scshape i\kern-0.25em b}\kern-0.8em\TeX}}}

\usepackage{multirow}
\usepackage{float}

\usepackage{enumitem}

\begin{document}
\title{Urban Mobility Assessment Using LLMs}

\author{Prabin Bhandari}
\email{pbhanda2@gmu.edu}
\orcid{0009-0006-9034-6372}
\affiliation{%
  \department{Department of Computer Science}
  \institution{George Mason University}
  \streetaddress{4400 University Dr}
  \city{Fairfax}
  \state{Virginia}
  \country{USA}
  \postcode{22030}
}

\author{Antonios Anastasopoulos}
\email{antonis@gmu.edu}
\affiliation{%
  \department{Department of Computer Science}
  \institution{George Mason University}
  \streetaddress{4400 University Dr}
  \city{Fairfax}
  \state{Virginia}
  \country{USA}
  \postcode{22030}
}

\author{Dieter Pfoser}
\email{dpfoser@gmu.edu}
\affiliation{%
  \department{Department of Geography and Geoinformation Science}
  \institution{George Mason University}
  \streetaddress{4400 University Dr}
  \city{Fairfax}
  \state{Virginia}
  \country{USA}
  \postcode{22030}
}

\begin{abstract}

Understanding urban mobility patterns and analyzing how people move around cities helps improve the overall quality of life and supports the development of more livable, efficient, and sustainable urban areas.
A challenging aspect of this work is the collection of mobility data by means of user tracking or travel surveys, given the associated privacy concerns, noncompliance, and high cost.
This work proposes an innovative AI-based approach for synthesizing travel surveys by prompting large language models (LLMs), aiming to leverage their vast amount of relevant background knowledge and text generation capabilities.
Our study evaluates the effectiveness of this approach across various U.S. metropolitan areas by comparing the results against existing survey data at different granularity levels.
These levels include (i) pattern level, which compares aggregated metrics like the average number of locations traveled and travel time, (ii) trip level, which focuses on comparing trips as whole units using transition probabilities, and (iii) activity chain level, which examines the sequence of locations visited by individuals.
Our work covers several proprietary and open-source LLMs, revealing that open-source base models like Llama-2, when fine-tuned on even a limited amount of actual data, can generate synthetic data that closely mimics the actual travel survey data, and as such provides an argument for using such data in mobility studies.

\end{abstract}



\maketitle

\section{Introduction}
\label{sec:intro}

Analyzing mobility patterns in urban areas has become a critical area of research given the population growth in (mega)cities and associated space and resource constraints.
Analyzing how people navigate cities is essential for a number of tasks that have the ultimate goal of improving the quality of life of residents and supporting the development of more livable, efficient, and sustainable urban areas. Given the right insight, city planners can design better transportation systems, reduce traffic congestion, reduce environmental impacts, and ensure equitable access to transportation for all residents. 

Mobility data is the cornerstone of this assessment. Yet collecting such data presents significant challenges.
Current methods include user tracking \cite{liu2009urban, bazzani2010statistical, tang2015uncovering} and travel surveys \cite{pucher2003socioeconomics, mattson2012travel}. Each has their own set of issues such as privacy concerns, participant non-compliance, and high cost. 
This paper proposes an innovative approach to the assessment of urban mobility using artificial intelligence and specifically large language models to tap into the existing collective wisdom and so to overcome these challenges.

Mobility data, which captures the movement of people, is foundational for urban mobility assessment. 
While the rise of 5G and IoT technologies promises a great avenue to collect mobility data using GPS tracking, this approach has significant issues.
Collecting GPS data raises privacy concerns, and even when the data is collected securely and ethically, many people are reluctant to opt-in to such programs. In addition, pure tracking data does not capture the context and purpose of a trip.
Here, an alternative method is conducting travel surveys.
These surveys collect information about an individual, their household, and use a travel diary, which captures a person's movements on a particular day.
Although different approaches exist, in general, surveys ask their participants to log their start and end time, start and end location, mode of travel, and purpose of the journey in a travel diary.
The quality of such surveys is often disputed given their low response rate and associated high cost, e.g., the US National Highway Travel Survey (NHTS) from 2017 has an overall response rate of only 15.6\%~\cite{westat2017nhts}. 

Those concerns have motivated researchers to explore simulation-based techniques to gather synthetic travel survey data~\cite{greaves2000simulating}.
In this context, our work leverages large language models~(LLM) to generate travel surveys, offering a promising alternative to traditional methods and enhancing urban mobility assessment.

LLMs have revolutionized the field of natural language processing and artificial intelligence, eliminating the need for task-specific models trained using vast amounts of labeled datasets.
With the advent of pre-training techniques, LLMs, which have a large number of parameters in them, can be pre-trained without any labeled data.
This pre-training allows LLMs to encode different types of knowledge within their parameters, effectively working as knowledge bases~\cite{petroni-etal-2019-language}.
LLMs also encode world knowledge and exhibit common sense reasoning capabilities, enabling them to understand and generate human-like text across various contexts.
Demonstrating this capability, \citet{brown2020language} showed that sufficiently scaled LLMs like GPT-3 can handle diverse downstream tasks just by passing a task description, with or without a few sample task examples, as context, a technique known as Prompting.
Moreover, advancements in prompting techniques~\cite{bhandari2023prompting} have made it possible to handle complex tasks including reasoning.

\emph{We hypothesize that since LLMs are trained on a vast amount of textual data, they could potentially encode travel-related data within their parameters.}
This suggests that simulating travel surveys using LLMs could present a promising alternative.
In our study, we prompt LLMs, including Llama-2~\cite{touvron2023llama}, Gemini-Pro~\cite{reid2024gemini}, and GPT-4~\cite{achiam2023gpt}, to simulate a travel survey.

In our approach, an LLM generates a travel survey by completing the travel diary entries as a survey participant.
To validate the results, we compare them to actual travel surveys, specifically the 2017 NHTS data.
The evaluation of the results happens at the (i) pattern level, i.e., aggregate metrics such as the number of locations visited or total travel time, (ii) trip level, i.e., comparing trips between location pairs using transition probabilities, and (iii) activity chain
level, i.e., comparing daily location sequences. 
Our evaluation covers different US Statistical Metropolitan areas and compares them to the 2017 NHTS data.
In another experiment, we also compare our LLM-based generation results to a Patterns-Of-Life, an agent-based model (ABM) simulation \cite{kim2020location}.

Our findings reveal that LLMs even without fine-tuning or alignment to better follow human instructions, encode travel-related information.
However, fine-tuning these LLMs even with a small subset of actual travel surveys enables them to better mimic the real-world surveys, surpassing the performance of existing simulation techniques.

The remainder of this paper is structured as follows.
Section~\ref{sec:related_work} reviews related work in the field of travel survey simulation for urban mobility assessment.
An overview of our proposed LLM-based travel survey simulation system is presented in Section~\ref{sec:methodology}.
Section~\ref{sec:evaluation} discusses the methodology used to assess the quality of our generated data.
Section~\ref{sec:experiments} details our experimental setup and results.
We then compare our system with an agent-based simulation approach in Section~\ref{sec:pof-comparision}.
Finally, Section~\ref{sec:conclusion} provides conclusions and directions for future work.
\section{Related Work}
\label{sec:related_work}

Researchers have explored various techniques to simulate travel surveys, aiming not to replace the instrument but to complement and extend collected travel survey data.

One of the first techniques was based on the Monte Carlo method \cite{von1951monte}.
\citet{greaves2000creating} proposed a method that first employed the Classification and Regression Tree method~\cite{breiman1984classification} to classify the households from an actual survey into clusters based on travel attributes.
Subsequently, within each cluster, the variability of trip attributes was captured as a probability distribution.
This approach was then applied to households within the location of interest, where Monte Carlo Sampling (MCS) was utilized to select values from the appropriate probability distribution.
Initially applied in Baton Rouge, Louisiana, this technique has been validated in various locations, including the Dallas-Fort Worth and Salt Lake City metropolitan area~\cite{stopher2004monte}, and Adelaide and Sydney in  Australia~\cite{pointer2004monte}.
A similar approach was adopted by \citet{mohammadian2010synthetic}, who used a synthesized population for the target location and employed neural networks to facilitate the transferability of clusters from actual survey data to the synthesized population.
Although this method accurately predicted trip rates, it failed to capture other trip characteristics like mode, departure time, and travel time. 
The authors attribute this to the method's inability to reflect contextual differences between locations.
In contrast, our approach mitigates this issue, as recent studies have demonstrated the ability of LLMs to discern between locations when provided with the proper context~\cite{bhandari2023large}.

The method described by \citet{greaves2000creating} generates each trip in a chain independently of prior trips, which is also by default handled by the autoregressive nature of LLMs that we employ. Going beyond, researchers have also tried simulating trips based on previous trips taken in the chain~\cite{janssens2004simulation, kitamura1997generation}. 
The main idea here is to generate weighted transition probabilities, as we have used in our evaluations.
However, these methods typically rely on second-order transition probabilities only,  whereas during our model fine-tuning, we incorporate higher-order transition probabilities.

Last, another technique involves simulating real-world location-based social network (LBSN) datasets, based on typical human behavior, also known as patterns of life~\cite{kim2019simulating, kim2020location}.
In this simulation, agents with needs similar to real individuals are tracked to generate LBSN datasets resembling actual human interactions.
One advantage of this technique is that it can generate precise location information, which our system currently lacks.
To evaluate our system against this `patterns of life' based simulation, we conduct a comparative analysis in Section~\ref{sec:pof-comparision}.
\section{System Overview}
\label{sec:methodology}

\begin{figure}[t]
    \centering
    \includegraphics[width=\linewidth]{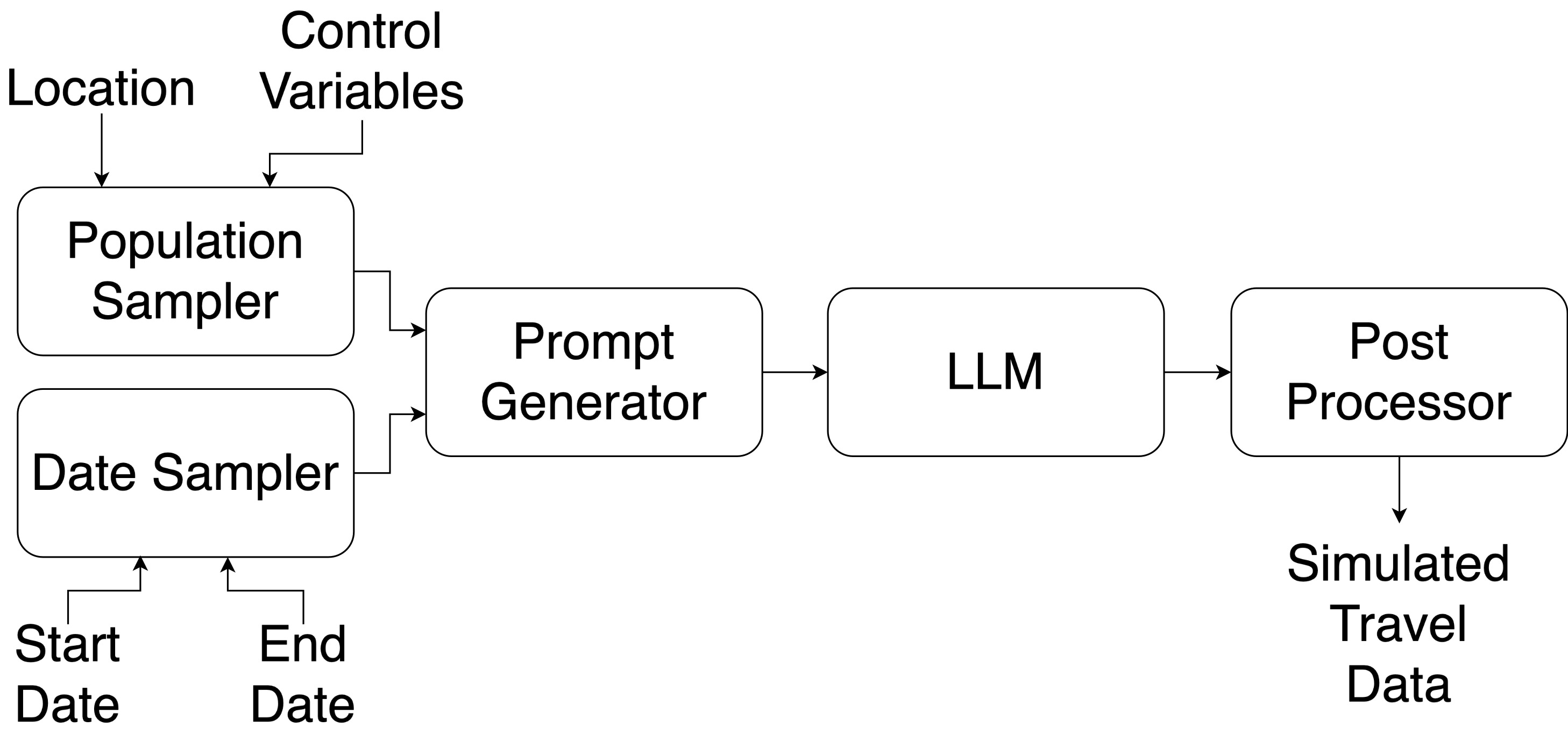}

    \caption{LLM-based travel survey generation system.}
    \Description{Overview of LLM-based travel generation system}
    \label{fig:system_overview}
\end{figure}

Our proposed LLM-based system aims to generate synthetic travel survey data to simplify and complement the data aspect of urban mobility assessment.
Leveraging the extensive knowledge encoded in LLMs, the system generates travel survey responses by prompting the LLM to fill out a travel diary. 
The system then processes the LLM-generated output, which is structured as a travel diary table. 
We show a sample table of the generated output in Table~\ref{tab:sample_output}
The overall system architecture is given in Figure~\ref{fig:system_overview}, comprising the following components: population sampler, date sampler, prompt generator, LLM, and a post-processor.

\begin{table}[t]
    \centering
    \caption{A (reformatted) sample output generated by our LLM-based approach.}
    \label{tab:sample_output}
    \begin{tabular}{c|c|c|c}
    \toprule
         Place &Arrival &Departure & Location \\
         Visited&Time&Time& Type \\
         \midrule
         Home & 00:00 AM & 07:30 AM & 1\\
         Work & 08:00 AM & 05:00 AM & 3\\
         Home & 05:30 AM & 11:59 PM & 1\\
         \bottomrule
    \end{tabular}
\end{table}


\noindent \textbf{Population Sampler} \ 
To ensure the accuracy of our LLM-based approach, it is important to account for the diverse demographics of the targeted location.
To achieve this, we employ a population sampler, sampling individuals based on various control variables such as sex, age group/age, race, school enrollment, participation in the labor force, employment, occupation, marital status, household type, and presence of children under 18 years old.
To maintain consistency with actual travel surveys, we exclude participants under the age of 16.
We use the American Community Survey (ACS)~\cite{bureau2021american} to define the prior probabilities used to sample demographics for locations within the United States of America.

\noindent {\textbf{Date Sampler}} \ 
Given that surveys operate within a specific duration, we employ a date sampler, which selects the survey date for each participant by uniformly sampling a date falling between the designated start and end dates of the survey.

\noindent {\textbf{Prompt Generator}} \ 
The prompt generator is responsible for crafting inputs for the LLM, using the outputs of the population and date samplers.
Our prompts are hand-designed to emulate the format of the National Household Travel Survey 2017 of the USA~\cite{federal_highway_administration_2017} employing travel diary-based completion prompts, aiming to simulate an individual's travel diary.
The prompt instructs the LLM to populate a table detailing the places visited by the participant, along with the arrival and departure times and the location types. Specifically, the pre-specified columns are: \texttt{`Place Visited'}, \texttt{`Arrival Time'}, \texttt{`Departure Time'}, and \texttt{`Location Type'}. 
An illustrative prompt is shown in Appendix~\ref{sec_a:prompt_templates} in Figure~\ref{fig:completion_prompt}.
We use the location type categorization used by the National Household Travel Survey 2017. We have provided the categorization in Appendix~\ref{sec_b:supporting_tables} in Table~\ref{tab:nhts_cat}.

\noindent {\textbf{Large Language Model}} \ 
We prompt the LLMs with the output generated by the prompt generator and assess their efficacy in simulating travel surveys using three LLMs: Llama-2, Gemini, and GPT-4.
In our Llama-2 setup, we use top-$k$ sampling-based decoding with $k=50$ and a temperature of 1. 
Top-$k$ sampling limits the token pool while decoding to the $k$ most likely options at each step, while temperature controls the randomness during token selection.
A higher temperature value increases randomness, while a lower temperature value makes the output more deterministic.
For Gemini and GPT-4, we use the default sampling-based decoding strategy offered in their Application programming interface~(API).

\noindent {\textbf{Post-processor}} \ 
While the prompt instructs the LLMs to produce output in a structured tabular format, their responses sometimes include extraneous text or omit required information.
Hence, we need to post-process the responses generated by the LLM to extract structured travel data. 

First, the post-processor utilizes regular expressions to extract the tabular data from the LLM-generated output.
However, it is important to note that not all survey outputs result in accurate table formation.
Specifically, for Llama-2 model outputs, the post-processor filters out surveys lacking travel time information or exhibiting travel times exceeding 2 hours between two locations.
In addition, specifically for Llama-2 model outputs, we exclude surveys featuring location types categorized as 97~(Other).
This decision is based on post-analysis, which revealed a notable prevalence of this location type, possibly due to the LLM trying to play it safe.

Last, the post-processor reclassifies the original 20 location types into a new set of 11 location types to facilitate an easier in-depth analysis.
The reclassification is provided in Table~\ref{tab:loc_types_reclassification}.

\begin{table}[t]
    \centering
    \begin{tabular}{l|l}
    \toprule
      NHTS-2017 Location Types   & New Location Types \\
    \midrule
        Regular home activities& Home\\
        Work from home& Home \\
        Work & Work \\
        Work-related meeting/trip & Work \\
        Volunteer activities & Community \\
        Drop off /pick up someone & In Transit \\
        Change type of transportation & In Transit \\
        Attend school as a student & Education\\
        Attend child care & Care \\
        Attend adult care & Care\\
        Buy goods  & Shopping \\
        Buy services  & Shopping \\
        Buy meals  & Eat Meal \\
        Other general errands  & Other \\
        Recreational activities  & Recreational \\
        Exercise  &  Recreational \\
        Visit friends or relatives & Social \\
        Health care visit  & Social \\
        Religious or other community activities & Community \\
        Something else & Other \\
    \bottomrule
    \end{tabular}
    \caption{Reclassification of the 20 location types in NHTS-2017 survey to 11 location types.}
    \label{tab:loc_types_reclassification}
\end{table}
\section{Evaluation Methodology}
\label{sec:evaluation}

Comparing our generated travel data to actual travel data is inherently complex and requires multiple metrics ranging from aggregate to more detailed levels, such as daily activity chains (cf. \citet{janssens2004simulation}).
These levels include \textbf{(i) pattern level}, which examines aggregated metrics such as the average number of locations traveled and average travel time, \textbf{(ii) trip level}, which focuses on comparing all the trips using transition probabilities, and \textbf{(iii) activity chain level}, which analyzes the sequence of locations visited by individuals.

\noindent {\textbf{Pattern level}} \ 
Pattern-level metrics compare the generated and actual travel data at the highest level by examining aggregate measures of travel behavior.
At this level, we use the average number of locations visited per participant and the average travel duration as key comparison metrics.
Additionally, we analyze location visit counts (visualized with histograms) and travel time distributions (visualized with box plots), comparing the characteristics of our generated data to those of actual travel survey data.

\noindent {\textbf{Trip level}} \ 
A trip refers to a movement or a journey made by an individual from one location to another (e.g., `Home -> Work').
Trips are the fundamental unit of analysis in urban mobility.
So, to compare trips, we generate transition probabilities, as shown in Table~\ref{tab:trans_probs_1}, from our actual and generated survey data. 
These transition probabilities capture the likelihood of traveling between different locations.
We also generate the second-order transition probabilities~(example provided in Table~\ref{tab:trans_probs} in Appendix~\ref{sec_b:supporting_tables}).
Effectively, the transition probabilities computed over a set of surveys can be presented as a square matrix. This representation allows us to easily quantify the distance between transition probabilities computed over different sets of surveys (actual and generated), by calculating the matrix norm of their difference (i.e., the norm of the subtraction --actual minus generated-- of the probability matrices).
In our case, we use the Forbenius norm~\cite{golub2013matrix}.
We additionally focus on \textit{destination} probabilities, i.e., the probabilities of a trip ending at a specific location type, to evaluate the likelihood of each location type across different LLMs and contrasting them with the destination probabilities observed in the actual survey data.

\begin{table}[ht]
\centering
\caption{Example of a first-order transition probabilities.}
\label{tab:trans_probs_1}
\begin{tabular}{l|rrrrr}
\toprule
\multirow{2}{*}{\(x_{t-1}\)} & \multicolumn{3}{c}{\(x_{t}\)} \\
& Home & Work & Community & In Transit & ... \\
\midrule
Home & 0 & 0.31 & 0.05 & 0.07 & ...  \\
Work & 0.21 & 0 & 0.03& 0.04 & ...  \\
... & ... & ... & ... & ... & ...  \\
\bottomrule
\end{tabular}
\end{table}

\noindent{\textbf{Activity chain level}} \ 
Activity chains represent the complete sequence of visited locations during a day (e.g., `Home -> Work -> Home').

We examine different combinations of the presence of activity chains across the actual and generated data.
First, for a given location we analyze if generated activity chains appear in the respective actual survey, i.e., evaluating for a specific location the portion of the survey data that is captured by the generated data. 
We term this metric \textit{`Chain Precision\textsubscript{loc}'} in our results.

Second, we consider the presence of generated chains within the actual chains across \textit{all} locations. The assumption here is that a specific activity chain is \textit{realistic} if it appears in an actual survey of \textit{any} location. 
In our tables, we call this metric \textit{`Chain Precision\textsubscript{all}'}.

The previous two metrics are conceptually equivalent to \textit{precision}. We test how many of our generated activity chains are realistic, i.e., similar to the ground truth. Thus, we also need a metric conceptually similar to \textit{recall}. We quantify the presence of activity chains from actual surveys in generated surveys, measuring the representativeness of the generated chains.
Specifically, we compute the ratio of the actual chains that are represented by the generated ones.
We name this metric as \textit{`Chain Recall'} in our tables.

Finally, we calculate the weighted overlap between the actual and generated activity chains.
This weighted overlap can serve as a single measure to compare different LLMs, indicating which LLM generates activity chains that most closely match the actual ratios.
Table~\ref{tab:activity_chain_measures} summarizes the different aspects of these activity chain level metrics for easy reference.

\begin{table}[h]
    \centering
    \caption{Activity-chain level metrics used to compare the generated and actual travel chains. Labels used in results and a short description.}
    \label{tab:activity_chain_measures}
    \begin{tabular}{l|p{5cm}}
    \toprule
      \textbf{Metric}   &  \textbf{Description}\\
    \midrule
    Chain Precision\textsubscript{loc} & Generated chain appears in respective actual survey (Resemblance of generated to actual) \\
    \midrule
    Chain Precision\textsubscript{all} & Generated chain appears in any actual (Presence in other locations also indicates realistic outputs)\\
    \midrule
    Chain Recall &  Actual chain appears in generated ones (Representatives of generated to actual )\\
    \midrule
    Weighted overlap & Similarity between actual \\
    between chains & and generated chains  \\
    \bottomrule
    \end{tabular}
\end{table}

In summary, assessing generated travel data against actual data requires a multi-level approach. 
We analyze aggregated metrics, individual trips, and activity chains.
These evaluations help us understand how generated data aligns with actual ones, providing insights into the use of LLMs for urban mobility assessment.
\section{Experiments}
\label{sec:experiments}

Our experiments evaluate the accuracy and, essentially, the viability of our LLM-based travel survey system for various metropolitan areas in the US.
We generate synthetic travel data for five different metropolitan areas and compare them to the actual 2017 National Household Travel Survey (NHTS) data.
We first introduce our experimental setup and provide a detailed analysis focusing on the San Francisco-Oakland-Hayward metropolitan area. 
Following this case study, we extend our experiments and discussion to four more metropolitan areas.
In a nutshell, the results reveal that our proposed approach is indeed viable. LLMs fine-tuned on even a limited amount of actual data can generate synthetic data that closely resemble actual survey data. 

\subsection{Experimental Setup}

We use our LLM-based travel survey generation system to generate travel data for five different metropolitan areas in the USA: (i) San Francisco-Oakland-Hayward, (ii) Los Angeles-Long Beach-Anaheim, (iii) Washington-Arlington-Alexandria, (iv) Dallas-Fort Worth-Arlington, and (v) Minneapolis-St. Paul-Bloomington.

The population and date samplers, the first two components of our system, are tasked with sampling for participant demographics and for dates to maintain consistency with actual travel surveys.
The population sampler uses the 2017 American Community Survey (5-year estimates) to sample individuals.
The date sampler samples a date between April 19, 2016, and April 25, 2017, to mimic the NHTS 2017 survey timeline.

LLM is a key component of our system that generates the generated travel data.
We use three different LLMs: (i) Gemini-Pro, (ii) GPT-4, and (iii) Llama-2,
to compare their effectiveness. 

Gemini-Pro and GPT-4-turbo are proprietary LLMs only accessible via an API.
We use the Gemini-Pro and GPT-4-turbo versions respectively.

Llama-2, an open-source LLM, is deployed on a GMU research computing GPU cluster.
We employ the largest variant of Llama-2, which has 70B parameters. However, we opt for the 8-bit quantized version, as fitting it in our GPU cluster is not feasible otherwise.  
Model quantization involves employing low-precision data types, such as 8-bit integers (int8), for model inference, reducing computational and memory demands, albeit with a minor performance trade-off.

The pre-trained Llama-2 model can be further fine-tuned to the task-at-hand.
Fine-tuning uses a subset of 10,000 randomly sampled surveys from NHTS-2017, including all possible locations.
These samples are then converted to produce prompts similar to ours based on the survey metadata (Figure~\ref{fig:completion_prompt} in Appendix~\ref{sec_a:prompt_templates}), while the LLM's output should align with the actual travel diary entries (Table~\ref{tab:sample_output}).
Fine-tuning the whole model is prohibitively expensive due to its size. We instead use Low-Rank Adaptation~\cite[LoRA;][]{hu2021lora}, which enhances pre-trained LLMs by freezing their model weights and injecting trainable rank decomposition within each layer of the LLMs transformer architecture.
This technique significantly reduces the number of trainable parameters, making fine-tuning LLMs feasible.
We employ AdamW~\cite{loshchilov2017decoupled} as the optimizer, with a learning rate of \(0.001\) and a cosine learning rate scheduler~\cite{loshchilov2016sgdr}, running for 3 epochs.
LoRA-based hyperparameters alpha and rank are set to 128 and 256 respectively.

In the following section, we present the results of employing these different LLMs, to provide an in-depth analysis of their performance for the San Francisco-Oakland-Hayward metropolitan area.
In a subsequent section, we use the insights from the detailed study to provide a comprehensive assessment for four more locations.

\subsection{Detailed evaluation - San Francisco}
\label{sec:intra-city-results}

What follows is a detailed analysis of the data generated by our LLM-based travel survey system for the San Francisco-Oakland-Hayward metropolitan area by comparing it to the actual 2017 NHTS data using the methodology outlined in Section~\ref{sec:evaluation}.

Our findings indicate the superior performance of the fine-tuned Llama-2 model, demonstrating its effectiveness even for locations / cities where no actual travel survey data is available for training the model.
Below we provide supporting arguments for all levels of analysis (pattern, trip, and activity chain). 
The final section also digs deeper into LLM fine-tuning to investigate the impact of training data choices and the model's generalization abilities.

\noindent \textbf{Pattern level evaluation} \ 
The pattern-level evaluation compares the actual and generated travel data at an aggregate level, i.e., in our cases consider the (average) number of visited locations and total travel time.

Our findings reveal that the Llama-2-trained model closely matches the actual data in terms of aggregate metrics of the average number of visited locations and travel time.

\begin{table}[t]
\centering
\caption{Comparison of the average number of visited locations and mean travel time across different models - Llama-2-trained model best matches survey data showcased by lower differences $\Delta$.}
\label{tab:intra_loc_travel_time}
\begin{tabular}{@{}l|rr@{ }cc@{ }c@{}}
\toprule
  &Num of& \multicolumn{2}{c}{Avg num of visited}  &  \multicolumn{2}{c}{Travel} \\
  &samples& locations & ($\Delta$) &  hours & ($\Delta$)\\
\midrule
Actual & 4027&5.35 &  & 1.74 & \\
\midrule
Gemini-Pro & 1241 &6.59 & (1.24) & 2.52 & (0.78) \\
GPT-4 & 688&6.96 & (1.61) & 2.52 & (0.78) \\
Llama-2 & 1279& \textbf{5.55} & \textbf{(0.20)} & 7.17 & (5.43) \\
\midrule
Llama-2-trained & 1452& \textbf{4.97} & \textbf{(-0.38)} & \textbf{1.53} & \textbf{(-0.21)} \\
\bottomrule
\end{tabular}
\end{table}

Table~\ref{tab:intra_loc_travel_time} presents the metrics for the various models and also indicates the respective sample sizes.
Among the four models, the Llama-2-trained model performs best since it produced, both, a reasonable average number of visited locations \textit{and} at the same time keeps travel time (in hours) close to the actual average. The other three models each perform poorly in at least one of the two metrics.

\begin{figure}[t]
    \centering
    \includegraphics[trim={1.3cm 1.3cm 1.2cm 1.2cm}, clip, width=\linewidth]{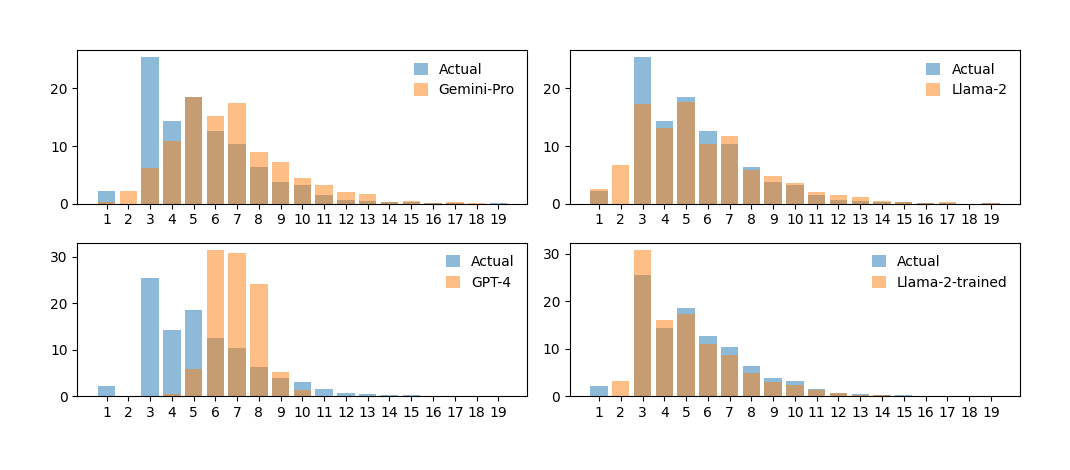}

    \caption{Distribution of the number of locations traveled by survey respondents for the San Francisco-Oakland-Hayward metropolitan area. Llama-2 and its fine-tuned variant show best alignment with actual survey data.}
    \Description{Distribution of the number of locations traveled by survey respondents for the San Francisco-Oakland-Hayward metropolitan area for different LLMs compared to the actual survey shows that Llama-2 and its fine-tuned variant are better aligned with the actual data.}
    \label{fig:hist_intra_city}
\end{figure}

Among the three base models, Llama-2 best matches the survey data in terms of visited locations but fails to properly capture travel time.
This discrepancy can be attributed to differences in model alignment. Model alignment refers to fine-tuning the model to better match the downstream task. GPT-4 and Gemini-Pro are fine-tuned for better alignment with human instructions, making them adept at handling numerical and temporal data. In contrast, the base Llama-2 model lacks such fine-tuning.
However, this alignment of GPT-4 and Gemini-Pro, while advantageous for travel time prediction, also leads to the generation of \textit{more text}, consequently resulting in an abnormally higher number of locations traveled.
This trend is also visible in Figure~\ref{fig:hist_intra_city} showing the distribution of the number of locations traveled by the survey respondents.
For Gemini-Pro and GPT-4, the number of locations traveled far exceeds the actual data, particularly for longer journeys. 
GPT-4 produces surveys with only 5-10 locations visited. Other models output ranges from 1 to 19 locations, which resemble more closely the distributions of the actual survey data.
The bottom right plot in Figure~\ref{fig:hist_intra_city} shows how very closely the fine-tuned Llama-2-trained model matches actual survey data in terms of locations visited.

\begin{figure}[t]
    \centering
    \includegraphics[trim={1.2cm 1.2cm 1.2cm 1.2cm}, clip, width=\linewidth]{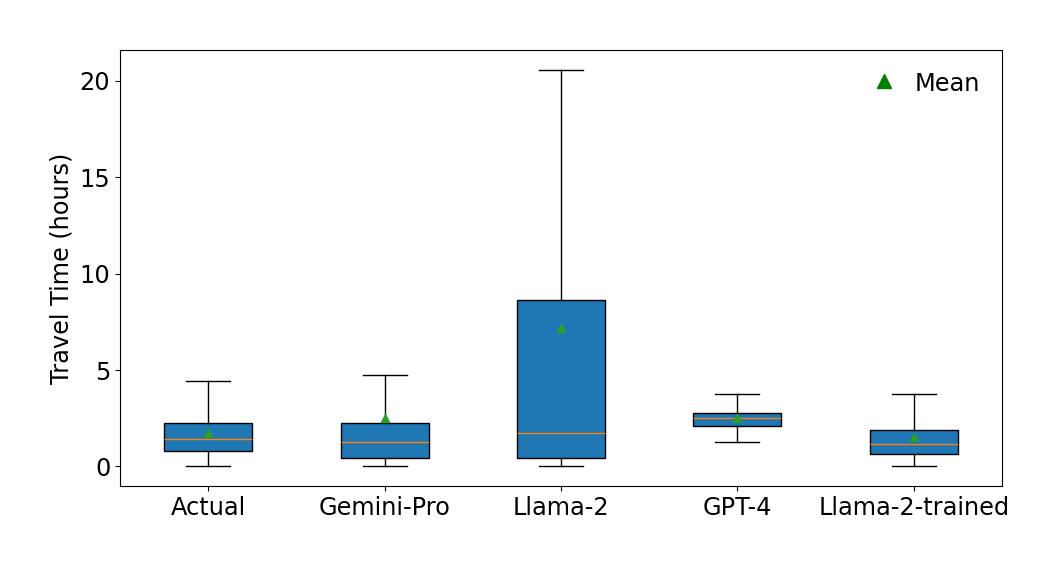}
    \caption{Travel time predictions - Gemini-Pro and GPT-4 models outperform Llama-2. The Llama-2-trained model matches survey travel times. Outliers are removed for better visualization.}
      \Description{Travel time predictions reveal that Gemini-Pro and GPT-4 models outperform Llama-2, while the fine-tuned Llama-2 model achieves similar accuracy. Outliers removed for better visualization.}
    \label{fig:tt_intra_city}
\end{figure}

The limitation of the Llama-2 base model in handling numerical data and consequently travel time is also evident in Figure~\ref{fig:tt_intra_city}, which shows the travel times of each model in a box plot.
The travel time for Llama-2 is as high as 120h (in a survey that captures 24h), indicating its struggle with time-related data.
Gemini-pro and GPT-4 better model travel times.

Fine-tuning the Llama-2 model addresses both of these issues. As shown in Figure~\ref{fig:tt_intra_city} (right-most box), the fine-tuned model's travel times more closely resembles those of the actual survey.

Overall, the pattern level evaluation of the various models indicates that the Llama-2-based generated data, particularly the \textit{Llama-2-trained model}, most closely matches the actual travel survey data. 

\noindent \textbf{Trip level evaluation} \ 
The trip level evaluation compares all the individual trips (between two locations) of the survey using transition probabilities. These results suggest that, again, the Llama-2-trained model best matches the travel survey transition probabilities.

\begin{table}[t]
\centering
\caption{Frobenius norm of transition probability differences between actual and generated data (lower is better). Llama-2-trained data best matches actual survey data.}
\label{tab:intra_norm2}
\begin{tabular}{l|rr}
\toprule
  & \multicolumn{2}{c}{Frobenius Norm with} \\
  & 1st order&  2nd order\\
  &transition prob &transition prob \\
\midrule
Gemini-Pro & 0.063 & 0.046 \\
GPT-4 & 0.166 & 0.130 \\
Llama-2 & 0.106 & 0.071 \\
\midrule
Llama-2-trained & \textbf{0.044} & \textbf{0.032} \\
\bottomrule
\end{tabular}
\end{table}

In the trip level comparison the Llama-2-trained model distinguishes itself. It shows the closest resemblance to the actual transition probabilities, as indicated by the Forbenius norm results of Table~\ref{tab:intra_norm2}.

\begin{figure*}[ht]
    \centering
    \includegraphics[trim={1.2cm 1.25cm 1.2cm 1.27cm}, clip, width=\linewidth]{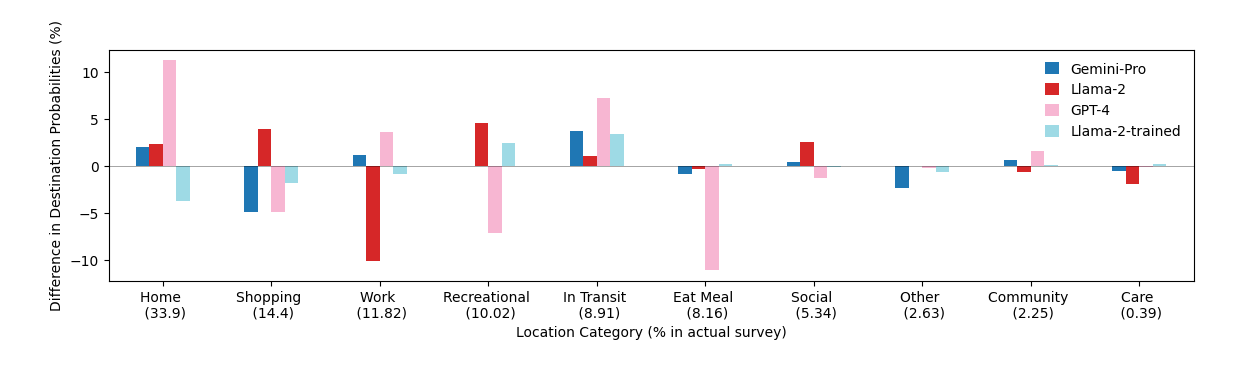}

    \caption{Difference in first order destination probabilities (actual - generated) per location category. Llama-2-trained outperforms (with shorter bars) other models.}
    \Description{Difference in first order destination probabilities (actual - generated).}
    \label{fig:dest_prob_intra_city}
\end{figure*}

Among the base models, Gemini-pro performs best, whereas GPT-4 shows the poorest alignment.
This variation in model performance can be further explained by examining the differences in first-order destination probabilities, which reflect the likelihood of a trip ending at a particular location type. 
Figure~\ref{fig:dest_prob_intra_city} shows the differences between the generated and actual survey data ordered by the probabilities of the actual survey.
Gemini-Pro and GPT-4 tend to overpredict destinations typically associated with tasks such as ``shopping'', ``dining out'', and ``recreational activities'' (notice the large negative bars).
This tendency is likely a result of the models being fine-tuned to respond to queries related to such location types.
Llama-2 tends to overpredict other location types, such as ``work''. However, it more closely captures the actual survey data trends. The fine-tuned Llama-2-trained model overall shows the best performance and alignment (as indicated by having generally smaller positive and negative bars in the figure).

The analysis of second-order destination probabilities detailed in the appendix shows a similar trend (see Figure~\ref{fig:dest_prob_2_intra_city} in Appendix~\ref{sec_c:supporting_figures}).

In summary, the trip-level evaluation indicates that Llama-2-trained generated data has transition probabilities best resembling those of the actual survey data. This finding is consistent with our pattern level evaluation.

\noindent \textbf{Activity chain level evaluation} \ 
The third type of evaluation is the most detailed one and compares the generated model results at the activity chain level by analyzing the sequence of locations visited during a 24h period.
Again, the results suggest that the Llama-2-trained model generates activity chains that more closely resemble the actual activity chains of travel surveys. 
This analysis shows that fine-tuning is crucial: we observe a 1.5-2x improvement in emulating the activity chains from travel surveys over the base Llama-2 model.

\begin{table*}[ht]
\centering
\caption{Activity chain analysis - Llama-2-trained model closely captures the activity chains of actual travel survey. Chain Precision\textsubscript{loc}: presence of generated chains within actual (resemblance); Chain Precision\textsubscript{all}: presence of generated chains within actual of all the cities (chain realism); Chain Recall: presence of actual chains within generated ones (Representativeness); Weighted overlap between chains: weighted similarity of chains.}
\label{tab:intra_chain_analysis}
\begin{tabular}{l|rr|rr|rr|r}
\toprule
  & \multicolumn{2}{c}{Chain Precision\textsubscript{loc}} & \multicolumn{2}{c}{Chain Precision\textsubscript{all}} & \multicolumn{2}{c}{Chain Recall} & Weight Overlap\\
\textit{high is better for all metrics}  & \% & weighted & \% & weighted &  \% & weighted & of Chains \\
\midrule
Gemini-Pro & 15.55 & 29.81 & 46.22 & 56.79 & 7.77 & 39.21 & 17.25  \\
GPT-4 & 7.89 & 11.92 & 40.71 & 52.76 & 1.63 & 2.19 & 1.44 \\
Llama-2 & 15.54 & 36.12 & 43.06 & 57.39 & 6.88 & 39.14 & 25.42\\
\midrule
Llama-2-trained & \textbf{30.88} & \textbf{61.16} & \textbf{65.44} & \textbf{80.10} & \textbf{11.92} & \textbf{48.22} & \textbf{42.54} \\
\bottomrule
\end{tabular}
\end{table*}

Table~\ref{tab:intra_chain_analysis} presents a comprehensive overview of the results for the San Francisco-Oakland-Hayward metropolitan area, showing all metrics following the explanation provided in Table~\ref{tab:activity_chain_measures}.

According to all metrics, Llama-2-trained outperforms all other models.
61\% of its generated activity chains appear in the actual survey (Chain Precision\textsubscript{loc}), compared to only around 36\% for the base Llama and even lower numbers for Gemini-pro and GPT-4.
Moreover, the Llama-2-trained model generates around 80\% of all (67\% unique) activity chains of the NHTS survey (see Chain Precision\textsubscript{all}). This is a significant improvement over the base models Llama-2, Gemini-pro, and GPT-4, which only reach 50\%.

Taking into account the frequency of the activity chains (by appropriate weighting), we find that approximately 47\% of the actual activity chains are present in the generated data of the Llama-2-trained model (see Chain Recall), compared to around 39\% for the Llama-2 base model and Gemini-pro, and only 2\% for GPT-4.
Finally, the weighted overlap between the actual and generated activity chains is around 42\% for the Llama-2-trained model, exceeding Gemini-pro by more than 20\% and GPT-4 by 40\%. This shows how good the Llama-2-trained model is at representing the activity chains present in the actual survey.

While matching already observed activity chains is a good indicator, it is important to remember that there are many possible activity chains. Hence, we need to also establish that the remaining generated but unmatched activity chains are still realistic.
To achieve this, we measure the correctness of unmatched activity chains using the Levenshtein distance~\cite{yujian2007normalized} (also known as \textit{edit distance}) and reporting the minimum Levenshtein distance between an unmatched activity chain and any actual one.
We find that among the remaining unmatched chains, 69\% have an edit distance of 1, 22.44\% have an edit distance of 2, 6.29\% have an edit distance of 4, and the remaining 2.75\% have a higher edit distance. 
This means that unmatched activity chains are not too dissimilar from those found in travel surveys.

\begin{figure}[t]
    \centering
    \includegraphics[trim={1.33cm 1.33cm 1.43cm 1.25cm}, clip, width=\linewidth]{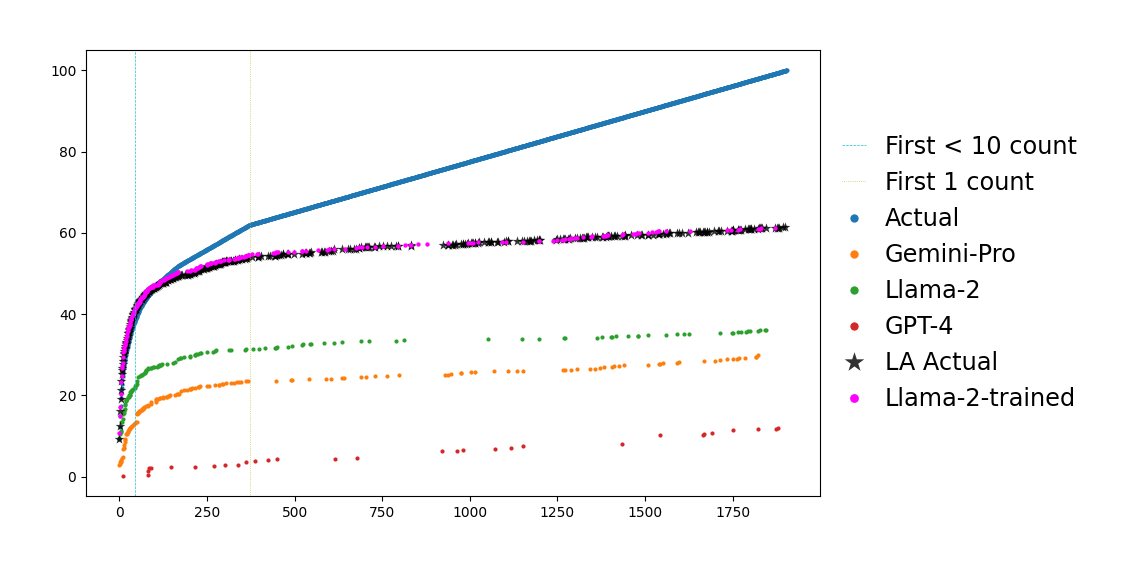}

    \caption{The cumulative sum of activity chain counts for the San Francisco-Oakland-Hayward metropolitan area's actual survey compared to the generated data. Llama-2-trained data closely matches the actual survey. Llama-2-trained and LA actual data are almost overlapping. }
    \Description{The cumulative sum of activity chain counts for the San Francisco-Oakland-Hayward metropolitan area's actual survey compared to the generated data. Llama-2-trained data closely matches the actual survey. Llama-2-trained and LA actual data are almost overlapping}
    \label{fig:chains_intra_city}
\end{figure}

Figure~\ref{fig:chains_intra_city} plots the cumulative sum of the activity chain counts in relation to the matched activity chain counts across various models, i.e., only generated activity that matches actual survey data are included. 
As a reference, we also plot the matching activity chains from another metropolitan area (Los Angeles-Long Beach-Anaheim~(LA)).
Two vertical lines in the plot indicate thresholds of activity chains with less than 10 counts (left blue line) and with only 1 count (right grey line). This shows the sparsity of many chains. 
From this figure, we can infer that GPT-4 exhibits the least resemblance, followed by Gemini-pro and then the Llama-2 base model (bottom three plots) to the actual data.
Finally, the Llama-2-trained model best matches the actual chains, akin to how the activity chains of the Los Angeles-Long Beach-Anaheim metropolitan area match the San Francisco-Oakland-Hayward metropolitan area data (overlapping plots of Llama-2-trained and LA data just below the actual survey data).
This result is intriguing, as we will see in the next sections that the difference between the travel surveys of different cities is very nominal, suggesting a noteworthy level of generation accuracy.

We also plot the distribution of the top 100 activity chains by the count of the actual survey in comparison to other generated surveys in Figure~\ref{fig:chain_hist_intra_city} in Appendix~\ref{sec_c:supporting_figures}.
These demonstrate that the Llama-2-trained results and LA's actual survey data closely match the distribution of San Francisco's activity chains. In contrast, GPT-4 matches only two chains, while Gemini and Llama-2 show better performance than GPT-4 but still fall short of Llama-2-trained.

Overall, activity chain level evaluation across various models indicates that the Llama-2-trained model data shows the closest resemblance to the activity chains present in the actual survey.

\noindent \textbf{Are location-specific training data important?} \
We have now established that fine-tuning an LLM to mimic travel surveys leads to producing realistic outputs that closely match real ones.
However, the data used to fine-tune the model can potentially play an important role in the model's performance.
Ideally, we would want to be able to generate travel surveys even for locations for which no real travel survey data are available -- this is in fact the ultimate frontier of synthetic travel data generation!

To test the ability of our generation approach to generalize beyond already-seen locations, we fine-tune another Llama-2 model, using a different fine-tuning dataset.
This dataset is curated by excluding the actual travel surveys from all five specific metropolitan areas where our system will be applied (including San Francisco). 

Encouragingly, our experiments show that the Llama-2-trained model without training data from the San Francisco-Oakland -Hayward area performs similarly to the model fine-tuned using such local data.

The original fine-tuned Llame-2-trained model that uses local data produces an average of 4.97 locations traveled and a travel time of 1.53h (as discussed previously in Table~\ref{tab:intra_loc_travel_time}). 
In comparison, the model fine-tuned without local data shows an average of 4.79 visited locations and a travel time of 1.60h. 
We consider these slight decrease in the number of locations and increase in travel time to be rather minor. 
The trip level evaluation also shows no discernible difference between the model fine-tuned with or without local data. The model fine-tuned with local data has a Forbenius norm of 0.044 for first-order transition probabilities, while the model without local data achieves a slightly better 0.043.
Conversely, for second-order transition probabilities, the model fine-tuned with local data has a Forbenius norm of 0.0319, compared to 0.0360 for the model fine-tuned without local data, indicating a slightly worse performance when lacking local training data.
The activity-chain level evaluation also indicates a minimal performance drop, with the weighted overlap between the actual and generated chains decreasing from 42.5\%  for the model fine-tuned with local data to 41\%  for the one fine-tuned without local data.

Overall, we find that a lack of local training data only very slightly decreases the quality of the generated data, and as such demonstrates the robust generalization capabilities of our LLM-based approach. 

In conclusion, the in-depth evaluation of our LLM-based travel survey generation system for the San Francisco area demonstrates that base models like Llama-2 
already produce meaningful data.
However, when fine-tuned even with a limited amount of data, these models can generate synthetic travel surveys resembling actual ones, even when location-specific training data is not available.

\subsection{Overall system performance}
\label{sec:inter-city-results}

Measuring the overall system performance of our travel survey generation system requires comparing results across multiple cities.
We have two desiderata for our system. First, that the generated data for the specific city match the actual travel surveys---this is in line with the analysis we performed on San Fransisco in \S\ref{sec:intra-city-results}, which we repeat here for all cities. And second, that generated surveys are \textit{meaningfully localized}: this means that they properly capture local urban infrastructure and population characteristics.
To this effect, we compare data generated for a specific city to the generated data for other cities, and we ground this discussion by comparing the actual travel survey data of different cities to each other. We also produce a clustering of the cities, the assumption being that more "similar" cities should also have more similar aggregate travel survey characteristics.

In short, we find that the Llama-2-trained model shows the smallest difference between generated and actual data among the same cities and across cities. 
Additionally, clustering cities shows that the Llama-2-trained model produces clusters that most closely resemble the actual city clusters, further validating the effectiveness of our approach. 

Table~\ref{tab:inter_loc_travel_time_chain_analysis} offers a consolidated pattern and activity chain level analysis for the additional four metropolitan areas. 
The table reaffirms our earlier finding that Llama-2-trained outperforms all other models. Of the base models, although it struggles with travel time, Llama-2 more closely approximates the actual data considering the pattern and activity chain level analyses.  

\begin{table}[t]
\centering
\caption{Pattern and activity chain level analysis for additional cities - Llama-2-trained produces results comparable to actual survey data.}
\label{tab:inter_loc_travel_time_chain_analysis}
\begin{tabular}{@{}l|c@{ }cc@{ }c@{}r@{}}
\toprule
  & \multicolumn{2}{c}{Avg. no} &   & & Wt. overlap \\
  & \multicolumn{2}{c}{of. locs} & \multicolumn{2}{c}{Travel} & of Actual \&\\
  & traveled & ($\Delta$) & hours & ($\Delta$)& Generated\\
\midrule 
&\multicolumn{5}{c}{Washington-Arlington-Alexandria} \\
Actual &5.15 & &1.75 && - \\
Gemini-Pro &6.45 &(1.30) & 2.82 &(1.07) &  18.89 \\
GPT-4 &6.91 &(1.76) & 2.58& (0.77) & 2.7\\
Llama-2 & \textbf{5.38} &\textbf{(0.23)} & 6.43 & (4.68) & 26.58\\
Llama-2-trained & 4.68 & (-0.47) & \textbf{1.56} & \textbf{(-0.19)} & \textbf{40.15}\\
\midrule
&\multicolumn{5}{c}{Los Angeles-Long Beach-Anaheim} \\
Actual &5.11 && 1.63 && - \\
Gemini-Pro &6.56 & (1.46) & 2.26 &(0.63) & 17.22 \\
GPT-4 & 6.88 &(1.77) & 2.53 &(0.90) & 1.51\\
Llama-2 & 5.63 &(0.52) & 6.22 &(4.59) & 27.29\\
Llama-2-trained & \textbf{4.64} &\textbf{(-0.47)} & \textbf{1.69} & \textbf{(0.06)} & \textbf{43.12}\\
\midrule
&\multicolumn{5}{c}{Dallas-Fort Worth-Arlington} \\
Actual &5.14 && 1.52 && - \\
Gemini-Pro &6.31 &(1.17) & 2.31 &(0.79) & 18.4 \\
GPT-4 & 6.91 &(1.77) & 2.53& (1.01) & 2.04\\
Llama-2 & \textbf{5.42}& \textbf{(0.28)} & 5.75 &(4.23) & 28.18\\
Llama-2-trained & 4.68 &(-0.45) & \textbf{1.56} &\textbf{(0.04)} & \textbf{44.42}\\
\midrule
&\multicolumn{5}{c}{Minneapolis-St. Paul-Bloomington } \\
Actual & 5.10 && 1.61 && - \\
Gemini-Pro &6.24 &(1.14) & 2.42 &(0.81) & 16.99\\
GPT-4 & 7.10 &(2.00) & 2.54 &(0.93) & 2.27\\
Llama-2 & 5.73 &(0.63) & 6.59 &(4.98) & 24.44\\
Llama-2-trained  & \textbf{4.75} &\textbf{(0.35)} & \textbf{1.44}& \textbf{(0.17)} & \textbf{45.84}\\
\bottomrule
\end{tabular}
\end{table}

\begin{figure*}[ht]
    \centering
    \includegraphics[trim={1.3cm 0cm 1.2cm 1.3cm}, clip, width=\linewidth]{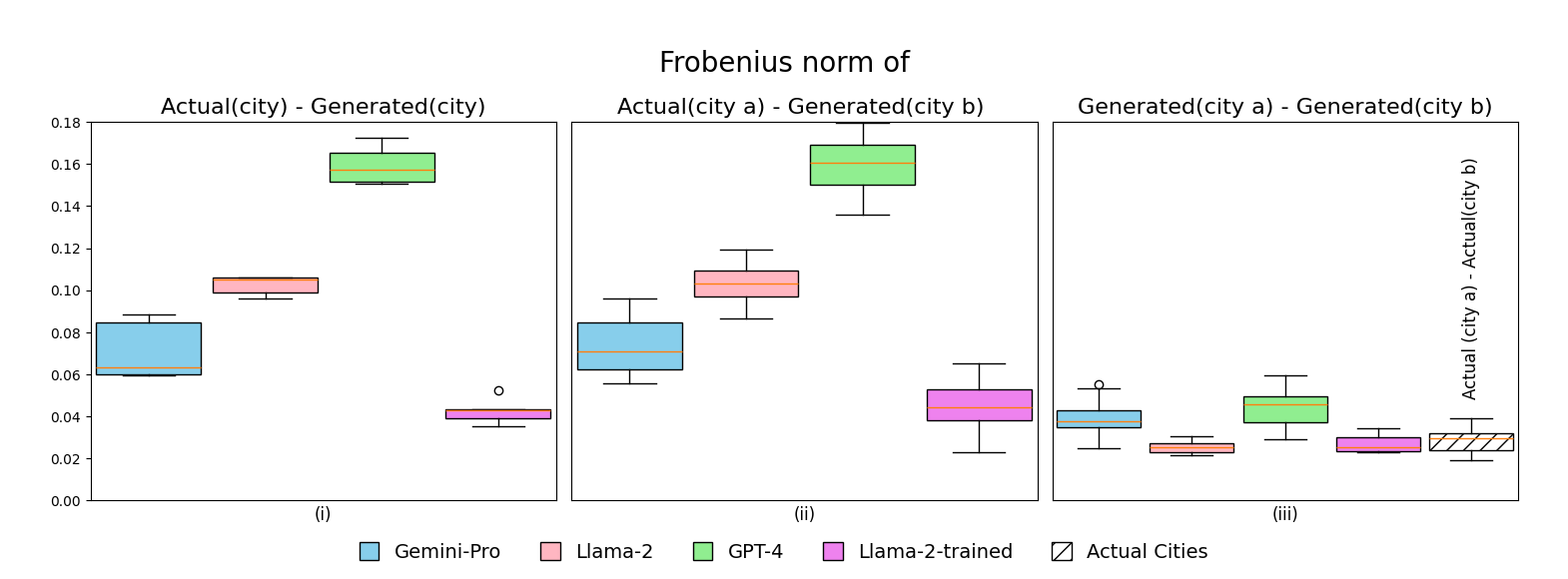}
    \caption{Forbenius norm for first-order transition probabilities of actual and generated surveys for different cities and for different LLMs. Minimal differences among actual cities themselves (see (iii): Actual) justify the minimal differences observed between actual and generated data of different cities (ii). Llama-2-trained shows best similarity (i). }
    \Description{Forbenius norm for first-order transition probabilities of actual and generated surveys for different cities in various combinations and for different LLMs. Minimal differences among actual cities themselves (see (iii): Actual) justify the minimal differences observed between actual and generated data of different cities (ii). Llama-2-trained shows best similarity (i). }
    \label{fig:norm2_inter_city}
\end{figure*}

The trip level analysis for different cities yields interesting insights.
Figure~\ref{fig:norm2_inter_city} presents box plots illustrating the Forbenius norm of the difference between the first-order transition probabilities of actual and generated surveys in various combinations for different LLMs.
Plot (i) compares the Forbenius norm for the difference of first-order transition probabilities of actual and generated surveys for the same city across different models, as in the previous analysis.
The distribution of norms is lowest for the Llama-2-trained model, which indicates that it best approximates the actual survey data.
Plot (ii) displays the norms between the actual and generated surveys for different cities. Effectively, we want to quantify how unique (or not) the generated surveys are to a specific city.
The plot is similar to Plot (i), albeit with slightly greater variability. 
This may seem counterintuitive, but the third plot sheds light on this phenomenon.
Plot (iii) shows the norm among generated surveys between different cities, together with the norm between actual surveys for different cities, i.e., how similar is the generated output of one model across different cities and how similar are actual surveys to each other (fifth shaded box in Plot (iii)).
The differences are overall much smaller than in Plots (i) and (ii), which means that the generated data of a model does not differ too much across different cities.
The same trend is observed for the actual data as well.
%
A takeaway from Figure~\ref{fig:norm2_inter_city} is that Llama-2-trained outperforms all other models and produces data that best matches actual survey data. 
Additionally, the similarity between Plots (i) and (ii) suggests that the difference among actual cities is also marginal, which is also shown in Plot (iii), indicating that the survey data for (US) cities are quite similar.

With the transition probabilities from actual and generated surveys from different LLMs, we also conduct Ward \textit{hierarchical clustering}~\cite{ward1963hierarchical}.
Using the \textit{actual} surveys, cities are clustered into two groups: 
\begin{enumerate}
    \item San Francisco-Oakland-Hayward, Los Angeles-Long Beach-Anaheim, and \textit{Washington-Arlington-Alexandria}
    \item Dallas-Fort Worth-Arlington and Minneapolis-St. Paul- \\Bloomington
\end{enumerate}

Using the generated data, clustering using surveys from the GPT-4 and Llama-2-trained models produces similar results. The only difference is Washington-Arlington-Alexandria moving to the other cluster (Dallas and Minneapolis).

This similarity of clusters, combined with Llama-2-trained generated data's similarity to the actual survey, continues to make a good case for using fine-tuned models in our travel survey generation approach. LLMs fine-tuned on even small amounts of data have the potential to serve as low-cost alternatives to generate travel survey data as part of an effort to simplify urban mobility assessment.


\section{Comparison to Patterns-of-Life Simulation}
\label{sec:pof-comparision}

The Patterns-Of-Life (POL) simulation for location-based social network datasets presents a novel approach to simulate travel surveys. The POL approach is based on an agent based modeling (ABM) framework that uses basic agent needs to generate mobility data for arbitrary geographic locations~\cite{kim2019simulating, kim2020location}.  
We compare the output of this approach to our LLM-based generation system using the pattern, trip, and activity level evaluation methodologies described in Section~\ref{sec:evaluation}. We are only using Llama-2-trained as it is our best performing LLM model. 

Our comparison uses the San Francisco-Oakland-Hayward Metropolitan area with the generated data generated by the Llama-2-trained model and the POL simulation. The latter produces continuous travel data for multiple agents, and to ensure consistency with travel surveys, we sample from all the data available, to match the same size of both simulation techniques.
One thing to note is that we could not extract the travel time for POL simulation, so we exclude the associated metrics from our analysis.
Another noteworthy consideration is the limited number of location types available in the POL simulation, which has only six different location types.
Consequently, we reclassify the original 20 different location types from NHTS-2017 into the 6 location types provided by POL (Reclassification shown in Table~\ref{tab:loc_types_reclassification_pof} in Appendix~\ref{sec_b:supporting_tables}) and modify our system's post-processor to allow for a direct, fair comparison.


\begin{table}[t]
\centering
\caption{Average number of locations traveled and the weighted overlap between actual and generated activity chains. Our LLM-based approach better aligns with actual data than the POL simulation.}
\label{tab:pof_loc_travel_time}
\begin{tabular}{l|rr@{ }cr}
\toprule
  && \multicolumn{2}{c}{Avg. no} & Weighted \\
  &No. of.& \multicolumn{2}{c}{of locations} & overlap\\
  &Samples& traveled & ($\Delta$)& of chains\\
\midrule
Actual & 4027&5.35 & & - \\
Llama-2-trained& 1435&4.98 & (-0.37) & \textbf{53.35}  \\
Pattern-of-life & 1435 & 3.98 & (-1.37) & 27.49\\
\bottomrule
\end{tabular}
\end{table}

\begin{figure}[t]
    \centering
    \includegraphics[trim={1.3cm 1.3cm 1.2cm 1.2cm}, clip, width=\linewidth]{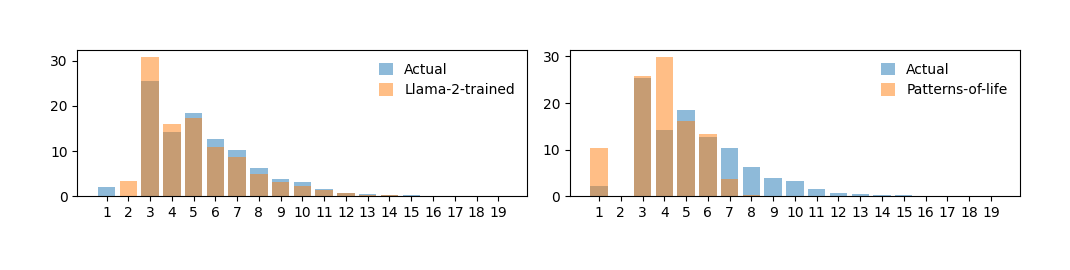}
    \caption{Distribution of the number of locations traveled. Our LLM-based approach better matches the actual distribution.}
    \Description{Distribution of the number of locations traveled reveals that our LLM-based approach better matches the actual~distribution.}
    \label{fig:hist_pof}
\end{figure}

The evaluations reveal some stark differences between the two approaches.
Our generative model better captures the average number of locations traveled by survey respondent as shown in Table~\ref{tab:pof_loc_travel_time}.
Figure~\ref{fig:hist_pof} shows the distribution of the number of locations. Both systems overpredict the number of visited locations, but the POL simulation has less variability and fails to capture the tail end of the distribution, which the LLM-based approach handles better.

The trip level evaluation again clearly shows the superiority of the LLM-based approach.
Table~\ref{tab:pof_norm2} provides the Forbenius Norm between the first-order and second-order transition probabilities obtained from the actual survey and both techniques.
Notably, our approach exhibits a significant, order-of-magnitude lower Forbenius norm than the POL simulation, indicating a substantial difference in their performance. 
The LLM-based approach significantly outperforms POL by a considerable margin in this metric.

\begin{table}[ht]
\centering
\caption{Frobenius Norm/transition probabilities comparison (lower is better). The LLM-based outputs more closely resemble trips of actual survey.}
\label{tab:pof_norm2}
\begin{tabular}{l|rr}
\toprule
  & \multicolumn{2}{c}{Frobenius Norm with} \\
  & 1st order&  2nd order\\
  &transition prob &transition prob \\
\midrule
Llama-2-trained & \textbf{0.037} & \textbf{0.034} \\
Patterns-of-life & 0.214 & 0.193 \\
\bottomrule
\end{tabular}
\end{table}

\begin{figure}[t]
    \centering
    \includegraphics[trim={1.3cm 1.35cm 1.2cm 1.28cm}, clip, width=\linewidth]{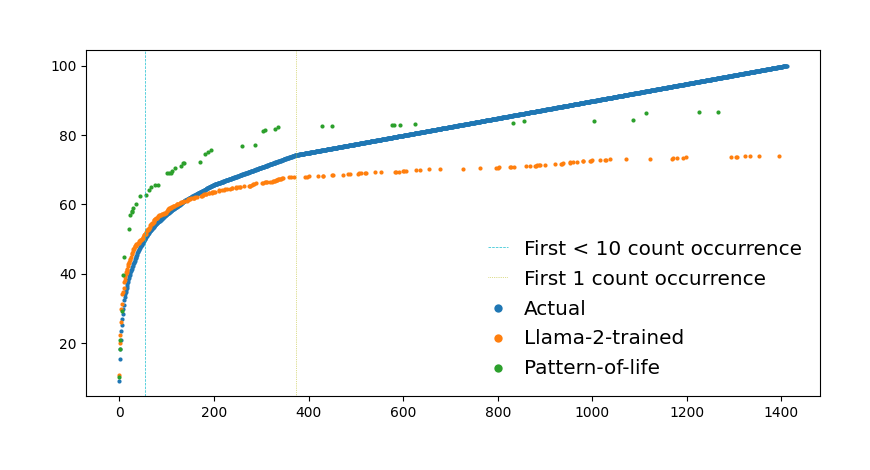}
    \caption{Cumulative sum of activity chain counts. POL simulation captures major patterns but fails to capture less common patterns, while our LLM-based system captures all patterns well.}
    \Description{Norm-2 between cities}
    \label{fig:chains_pof}
\end{figure}

The activity chain level evaluation, with the results shown in Table~\ref{tab:pof_loc_travel_time}, focuses on the weighted overlap of activity chains generated by the different methods.
The results in predicting activity chains as shown in Figure~\ref{fig:chains_pof}, where the cumulative sum of the activity chain counts is plotted compared to the matched activity chain counts across both approaches.
While the POL approach accurately models the most prevalent activity chains of the actual survey (towards the left of the $x$-axis), it tends to overpredict them.
Moreover, it does not capture the less common activity chains of the actual survey. This is shown by the scarcity of single-count chains in the POL result compared to the LLM approach.

In conclusion, the survey, trip, and activity chain level analysis shows that our LLM-based travel survey simulation approach more accurately reflects the actual survey data compared to POL simulation.
While the POL simulation certainly has its advantages compared to our technique (like generating precise location data), combining both techniques could lead to an even more robust mobility simulation.
We leave this as a future work.
\section{Conclusion and Future Work}
\label{sec:conclusion}

We propose a novel approach for urban mobility assessment utilizing Large Language Models~(LLMs).
Current methods of urban mobility data collection, a cornerstone for urban mobility assessment, often face significant challenges, including privacy concerns, noncompliance, and high costs.
To address these issues, we introduce an LLM-based travel survey generation system that generates synthetic travel data to facilitate enhanced urban mobility assessment.
We devise a set of evaluation criteria to assess the quality of generated data at pattern, trip, and activity chain levels.
Our findings highlight the effectiveness of our system, particularly when using base models like Llama-2, which has not been fine-tuned for better following human instruction but is trained with even a limited amount of actual travel data.

Future work will focus on extending this approach to locations outside of the USA and refining the system to generate more precise location data.
This will enhance the transferability and applicability of our system for global urban mobility assessment.

\begin{acks}
This work was supported by the National Science Foundation (Award 2127901) and by the Intelligence Advanced Research Projects Activity (IARPA) via Department of Interior/ Interior Business Center (DOI/IBC) contract number 140D0419C0050.  The U.S. Government is authorized to reproduce and distribute reprints for Governmental purposes notwithstanding any copyright annotation thereon. Disclaimer: The views and conclusions contained herein are those of the authors and should not be interpreted as necessarily representing the official policies or endorsements, either expressed or implied, of IARPA, DOI/IBC, or the U.S. Government. Additionally, this work was supported by resources provided by the Office of Research Computing, George Mason University and by the National Science Foundation (Award Numbers 1625039, 2018631).
\end{acks}

\bibliographystyle{ACM-Reference-Format}
\bibliography{sample-base}

\clearpage
\appendix

\section{Prompt Template}
\label{sec_a:prompt_templates}
\setcounter{figure}{0}
\renewcommand{\thefigure}{\Alph{section}\arabic{figure}}
\begin{figure}[H]
\caption{An example of a travel diary-based prompt.}
\centering\noindent\fbox{%
   \footnotesize \tt \parbox{0.95\linewidth}{%
{\vspace{0.5em}
The individual is a 59-year-old female whose racial background is `White alone'. Currently, she is not enrolled in school and is participating in the labor force. She is employed and working in the `Business and financial operations occupations' field. Regarding her marital status, she is married, and lives in a married couple family.  She lives in San Francisco, CA. She has been selected for a travel survey and has recorded her travel logs for 2016-05-05 which is a Thursday. 
She was asked to provide a list of all the places she visited on her travel date, including the exact times of arrival and departure, and the location type. The table format provided was as follows: \\

| Place Visited           | Arrival Time    | Departure Time  | Location Type   |\\
|---------------|--------------|----------------|---------------|\\
| [Place Name]            | [HH:MM AM/PM]   | [HH:MM AM/PM]   | [Location Type] |\\
| [Place Name]            | [HH:MM AM/PM]   | [HH:MM AM/PM]   | [Location Type] |\\
| ...                     | ...             | ...             | ... |\\

She was instructed to fill in each row with the relevant information for each place she visited on the specified date. If she visited the same place multiple times on the same date, she was advised to add a separate row for each visit to that place. \\

She was reminded of the following:\\
\\
1. Ensure that `Home' is included in the list if it was part of travel activities on the specified date.\\
2. She was asked to input the exact arrival and departure time as she noted in her travel diary.\\
3. She was advised to carefully enter the times, as gaps between the departure time of the previous place and the arrival time of the current place indicate the time taken to arrive at the current location.\\

For the `Location Type,' she was instructed to use the corresponding code from the provided list:

1: Regular home activities (chores, sleep)\\
2: Work from home (paid)\\
3: Work\\
4: Work-related meeting / trip\\
5: Volunteer activities (not paid)\\
6: Drop off / pick up someone\\
7: Change type of transportation\\
8: Attend school as a student\\
9: Attend child care\\
10: Attend adult care\\
11: Buy goods (groceries, clothes, appliances, gas)\\
12: Buy services (dry cleaners, banking, service a car, etc)\\
13: Buy meals (go out for a meal, snack, carry-out)\\
14: Other general errands (post office, library)\\
15: Recreational activities (visit parks, movies, bars, etc)\\
16: Exercise (go for a jog, walk, walk the dog, go to the gym, etc)\\
17: Visit friends or relatives\\
18: Health care visit (medical, dental, therapy)\\
19: Religious or other community activities\\
97: Something else\\

The table she created is as follows:

}
}}
\label{fig:completion_prompt}
\end{figure}

\section{Supporting Figures}
\label{sec_c:supporting_figures}
\setcounter{figure}{0}
\renewcommand{\thefigure}{\Alph{section}\arabic{figure}}

\begin{figure}[H]
    \centering
    \includegraphics[trim={1.2cm 1.25cm 1.2cm 1.27cm}, clip, width=\linewidth]{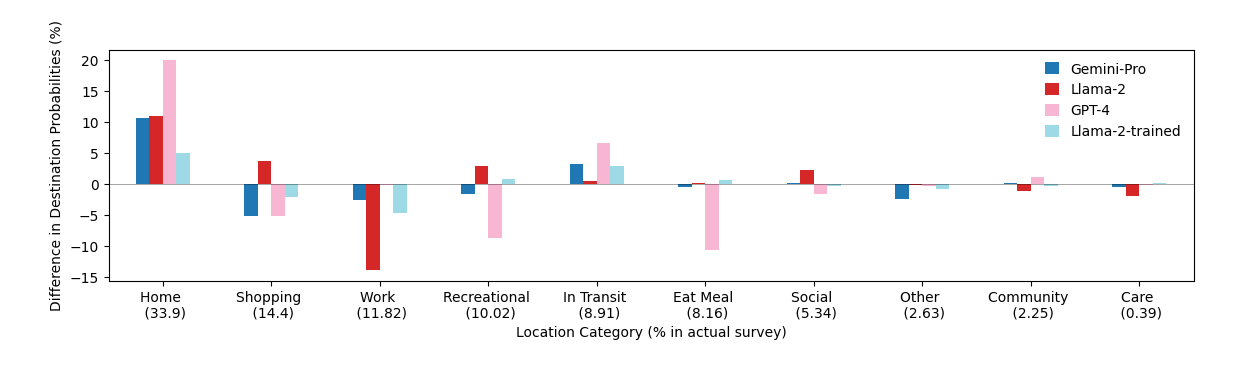}

    \caption{Difference in second-order destination probabilities (original - generated) for San Francisco-Oakland-Hayward metropolitan area .}
    \Description{Difference in second-order destination probabilities (original - generated).}
    \label{fig:dest_prob_2_intra_city}
\end{figure}

\begin{figure}[H]
    \centering
    \includegraphics[trim={1.35cm 1.33cm 1.30cm 1.33cm}, clip, width=\linewidth]{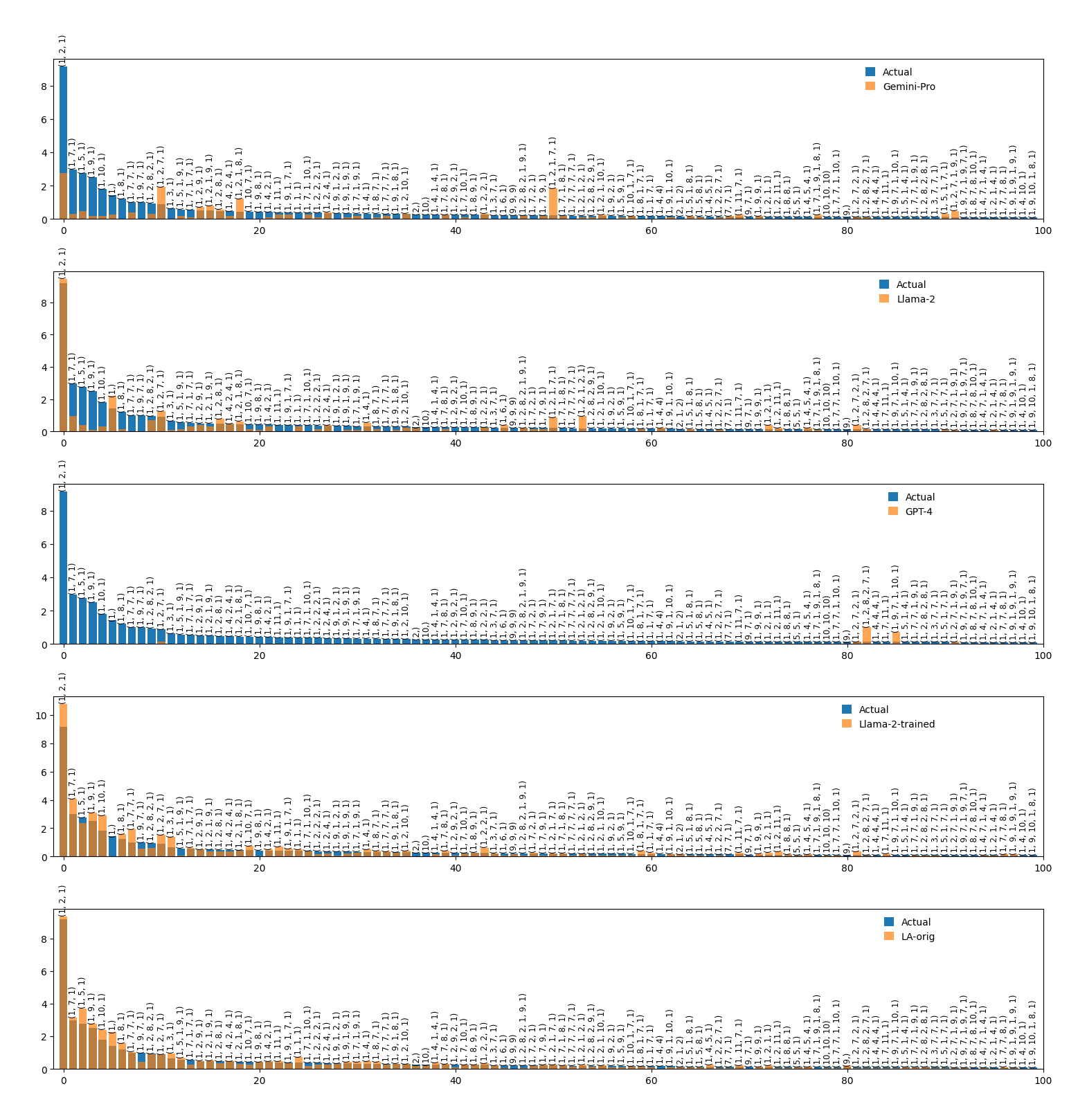}
    \caption{Distribution of top 100 activity chains of San Francisco-Oakland-Hayward metropolitan area in comparison to various LLM-based simulation techniques.}
    \Description{Distribution of top 100 activity chains of San Francisco-Oakland-Hayward metropolitan area in comparison to various LLM-based simulation techniques.}
    \label{fig:chain_hist_intra_city}
\end{figure}

\section{Supporting Tables}
\label{sec_b:supporting_tables}
\setcounter{table}{0}
\renewcommand{\thetable}{\Alph{section}\arabic{table}}

\begin{table}[H]
\caption{Categorization of location types done in the National Highway Travel Survey 2017.}
\begin{center}
\begin{tabular}{rl}
\toprule
  \multicolumn{1}{c}{\bf ID } & \multicolumn{1}{c}{\bf Label} \\
\midrule
1 & Regular home activities (chores, sleep) \\
2 & Work from home (paid)\\
3 & Work \\
4 & Work-related meeting / trip \\
5 & Volunteer activities (not paid) \\
6 & Drop off /pick up someone \\
7 & Change type of transportation \\
8 & Attend school as a student \\
9 & Attend child care \\
10 & Attend adult care \\
11 & Buy goods (groceries, clothes, appliances, gas) \\
12 & Buy services (dry cleaners, banking, service a car, etc) \\
13 & Buy meals (go out for a meal, snack, carry-out) \\
14 & Other general errands (post office, library) \\
15 & Recreational activities (visit parks, movies, bars, movies, etc) \\
16 & Exercise (go for a jog, walk, walk the dog, go to the gym, etc) \\
17 & Visit friends or relatives \\
18 & Health care visit (medical, dental, therapy) \\
19 & Religious or other community activities \\
97 & Something else \\
\bottomrule
\end{tabular}
\end{center}
\label{tab:nhts_cat}
\end{table}

\begin{table}[H]
\caption{Example of second order transition probabilities.}
\begin{center}
\begin{tabular}{l|l|rrr}
\toprule
\multirow{2}{*}{\(x_{t-2}\)} & \multirow{2}{*}{\(x_{t-1}\)}& \multicolumn{3}{c}{\(x_{t}\)} \\
& & Home & Work & ... \\
\midrule
Home & Work & 0.15 & 0.03 & ...  \\
Home & Community & 0.05 & 0.04 & ...  \\
... & ... & ... & ... & ... \\
\bottomrule
\end{tabular}
\end{center}
\label{tab:trans_probs}
\end{table}


\begin{table}[H]
    \caption{Reclassification of the 20 location types in NHTS-2017 survey to 6 location types of patterns of life based simulation system.}
    \centering
    \begin{tabular}{l|l}
    \toprule
      NHTS-2017 & Pattern of Life \\
    \midrule
        Regular home activities& Home\\
        Work from home& Home \\
        Work & Work \\
        Work-related meeting/trip & Work \\
        Volunteer activities & Recreation \\
        Drop off /pick up someone & Other \\
        Change type of transportation & Other \\
        Attend school as a student & School \\
        Attend child care & Other \\
        Attend adult care & Other \\
        Buy goods  & Other \\
        Buy services  & Other \\
        Buy meals  & Restaurant \\
        Other general errands  & Other \\
        Recreational activities  & Recreation \\
        Exercise  &  Recreation \\
        Visit friends or relatives & Other \\
        Health care visit  & Other \\
        Religious or other community activities & Other \\
        Something else & Other \\
    \bottomrule
    \end{tabular}
    \label{tab:loc_types_reclassification_pof}
\end{table}

\end{document}